# Revolutionizing Single Cell Analysis: The Power of Large Language Models for Cell Type Annotation

Zehua Zeng [ab], Hongwu Du [a]*

[a] Daxing Research Institute, University of Science and Technology Beijing, Beijing 100083, China.

[b] School of Chemistry and Biological Engineering, University of Science and Technology Beijing, Beijing 100083, China

Address: 112 Lab, Lihua BLDG, No.30 Xueyuan Road, Haidian District, Beijing 100083 China. Prof. Hongwu Du,

Email: hongwudu@ustb.edu.cn

## Abstract

In recent years, single cell RNA sequencing has become a widely used technique to study cellular diversity and function. However, accurately annotating cell types from single cell data has been a challenging task, as it requires extensive knowledge of cell biology and gene function. The emergence of large language models such as ChatGPT and New Bing in 2023 has revolutionized this process by integrating the scientific literature and providing accurate annotations of cell types. This breakthrough enables researchers to conduct literature reviews more efficiently and accurately, and can potentially uncover new insights into cell type annotation. By using ChatGPT to annotate single cell data, we can relate rare cell type to their function and reveal specific differentiation trajectories of cell subtypes that were previously overlooked. This can have important applications in understanding cancer progression, mammalian development, and stem cell differentiation, and can potentially lead to the discovery of key cells that interrupt the differentiation pathway and solve key problems in the life sciences. Overall, the future of cell type annotation in single cell data looks promising and the Large Language model will be an important milestone in the history of single cell analysis.

## Maintext

Cell type annotation has evolved significantly in 2023, with a variety of approaches being used. These include traditional manual annotation of literature readings, manual annotation of databases, automated annotation by machine learning, and migration annotation of datasets[1]. Single cell type annotation algorithms have typically been optimized for annotation accuracy and efficiency, with recent algorithms like scBERT demonstrating superior accuracy over previous methods[2]. However, such models have limited generalization capability and cannot handle cell or species types that are not present in the training data[3].

To annotate rare cell types and functional subtypes of cells, researchers still rely on clusters of corresponding marker genes, which are then paired with literature research and their own knowledge to complete the annotation[3]. This process is time-consuming and requires significant effort, and the accuracy still needs to be assessed by multiple experts. However, the emergence of large language models, such as ChatGPT, may render manual cell type annotation a thing of the past.

Previous studies have used self-clustering algorithms, such as unsupervised clustering, Dirichlet latent semantic inference, and non-negative matrix decomposition, to quickly group individual cells into different clusters[4]. These clusters are then annotated using existing cell type dictionaries to identify broad classes of cells, such as B cells and T cells, based on characteristic genes like MS4A1, CD79A, CD3D, and CD3E[5]. Marker genes specific to different clusters are then identified using a differential expression algorithm, allowing for more detailed annotation. However, due to the heterogeneity between clusters, marker genes are not always fixed, and large marker databases like Cellmarker and PanglaoDB have limitations, such as not including relevant species or cell types in your data[6,7].

To accurately annotate rare cell types, rare tissues, and cell subtypes, it is often necessary to manually infer the corresponding cell type based on marker gene function and the annotator's background knowledge[1]. This process can be time-consuming and requires a deep understanding of the relevant literature. For example, in one study, the marker genes found in the Cellmarker database corresponded to germ cells, despite the source of the single cell data being skin tissue[6]. Therefore, it is important to carefully document marker genes and enrich them for function to ensure accurate annotation.

The emergence of large language models, like ChatGPT and New Bing, in 2023 has provided a nearly perfect solution to the cell type annotation problem (Figure.1). Due to their vast training datasets, which include extensive scientific literature, these models are capable of answering a wide range of scientific questions, such as those related to the evolutionary history of fish, the development of new drugs, and the comparison of gene functions across different species.

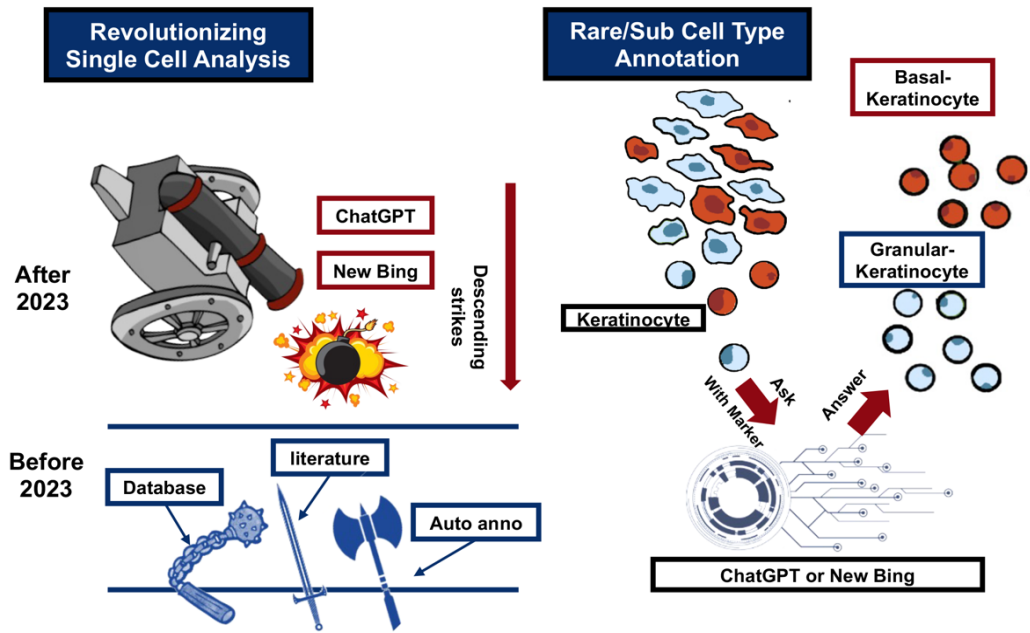

Figure 1 Revolutionizing Single Cell Analysis (left panel), A novel annotation method by Large language model (right panel).

Despite the vast potential of these language models, some members of the academic community have expressed concerns about their use in literature reviews, suggesting that they may negatively impact scientific work[8]. However, this viewpoint overlooks the fact that language models like ChatGPT are precisely designed to integrate the literature behind a literature review, making them an ideal tool for cell type annotation. With the help of these models, the time and effort required for manual annotation can be greatly reduced, while still ensuring that the resulting annotations are both accurate and reliable[9].

To illustrate the utility of large language models in cell type annotation, let's consider an example of single cell data analysis from mouse skin cells. In this study, we used an existing cell type dictionary to identify two broad categories of cells, fibroblasts and keratinocytes. However, as we were interested in understanding which cells in the skin were affected by a particular drug we were studying, we needed to further classify the keratinocytes into more specific subtypes.

To do this, we employed a self-clustering algorithm and conducted differential expression analysis to obtain a list of marker genes specific to each cluster. However, if a researcher had limited knowledge of mouse stratum corneum and was only familiar with the human counterpart, they would have required extensive literature searches to infer and annotate the functions of the different genes and to identify the specific subtypes of keratinocytes (KCs) (such as basal KCs, granular layer KCs, and spinous layer KCs)[10]. With the help of a large language model, like ChatGPT, these time-consuming and laborious tasks could be streamlined, enabling researchers to more efficiently and accurately annotate the different subtypes of cells and further our understanding of their functions.

With the help of large language models like ChatGPT, annotating cell subtypes or rare cells has become much more efficient and accurate. Instead of manually sifting through a vast amount of literature to determine the function of a handful of genes, researchers can simply input 3-5 marker genes of cluster and the background knowledge that the cluster is a keratinocytes into ChatGPT. The language model will then integrate all of the relevant literature it has analyzed and provide information on the functions of the input genes in keratinocytes, as well as infer the possible cell type of the cluster based on common features. This breakthrough greatly streamlines the annotation process, improving both efficiency and accuracy (Figure.1) (Table.1).

Table 1 Examples of questions for Rare/Sub cell type annotation

| Question: | Hi, I would like to know when Keratinocyte expressed Krt25 and Krt28 highly, what happen? |
|---|---|
| Answer (ChatGPT): | Krt25 and Krt28 are both keratin proteins that are primarily expressed in the skin's epidermal layer, specifically in the basal and suprabasal layers of the epidermis. These proteins are part of the keratinocyte differentiation program, which is a complex process that allows keratinocytes to develop from basal stem cells to fully differentiated, functional skin cells.<br><br>When Krt25 and Krt28 are highly expressed in keratinocytes, it generally indicates that these cells are in a more advanced stage of differentiation. This can occur during normal skin development, wound healing, or in response to environmental stresses such as UV radiation.<br><br>High expression of Krt25 and Krt28 is often associated with a stronger skin barrier function, as these proteins contribute to the structural integrity of the epidermis. However, it is important to note that the expression of these proteins is just one aspect of the complex process of keratinocyte differentiation, and additional factors must also be taken into account when interpreting the significance of their expression levels. |

In the near future, we can envision a world where any researcher can swiftly annotate single cell data and obtain accurate cell types. This breakthrough in cell type annotation will enable researchers to correctly analyze the function of cells[4]. Through the proposed chronological analysis, we can link trajectories to function, and uncover specific differentiation pathways of cell subtypes that were previously neglected. Moreover, with the assistance of ChatGPT or more large language models, we can obtain cells aligned with these specific trajectories. This will allow us to gain a more comprehensive understanding of cell trajectories, particularly in the study of cancer invasion, mammalian development, and stem cell differentiation[11-14]. By tracing cell differentiation trajectories, we can pinpoint key cells that disrupt the differentiation pathway and address major problems in life sciences. For instance, we can block intermediate cells in cancer progression, maintain stem cell stemness, and identify hubs in embryonic development. Overall, this breakthrough in cell type annotation represents a significant milestone in the history of single cell analysis.

# Acknowledgments

This work was financially supported by the Fundamental Research Funds for the Central Universities (FRF-TP-22-007A1), the Hebei Provincial Department of Science and Technology (No.19942410G), and the Student Research Training Program of University of Science and Technology Beijing.

# Reference

1   Pasquini, G., Arias, J. E. R., Schäfer, P. & Busskamp, V. Automated methods for cell type annotation on scRNA-seq data. *Computational and Structural Biotechnology Journal* **19**, 961-969 (2021).

2   Yang, F. *et al.* scBERT as a large-scale pretrained deep language model for cell type annotation of single-cell RNA-seq data. *Nature Machine Intelligence* **4**, 852-866 (2022).

3   Kharchenko, P. V. The triumphs and limitations of computational methods for scRNA-seq. *Nature Methods* **18**, 723-732 (2021).

4   Kiselev, V. Y., Andrews, T. S. & Hemberg, M. Challenges in unsupervised clustering of single-cell RNA-seq data. *Nature Reviews Genetics* **20**, 273-282 (2019).

5   Yang, S. *et al.* Decontamination of ambient RNA in single-cell RNA-seq with DecontX. *Genome biology* **21**, 1-15 (2020).

6   Zhang, X. *et al.* CellMarker: a manually curated resource of cell markers in human and mouse. *Nucleic acids research* **47**, D721-D728 (2019).

7   Franzén, O., Gan, L.-M. & Björkegren, J. L. PanglaoDB: a web server for exploration of mouse and human single-cell RNA sequencing data. *Database* **2019** (2019).

8   Aydın, Ö. & Karaarslan, E. OpenAI ChatGPT generated literature review: Digital twin in healthcare. *Available at SSRN 4308687* (2022).

9   Wang, Z., Ding, H. & Zou, Q. Identifying cell types to interpret scRNA-seq data: how, why and more possibilities. *Briefings in functional genomics* **19**, 286-291 (2020).

10  Norris, D. A., Shellman, Y. & Bellus, G. A. Keratinocytes. *Apoptosis and Inflammation*, 121-147 (1999).

11  Guo, Q. *et al.* CellTracer: a comprehensive database to dissect the causative multilevel interplay contributing to cell development trajectories. *Nucleic Acids Research* **51**, D861-D869 (2023).

12  Li, L. *et al.* What are the applications of single-cell RNA sequencing in cancer research: a systematic review. *Journal of Experimental & Clinical Cancer Research* **40**, 1-12 (2021).

13  Meistermann, D. *et al.* Integrated pseudotime analysis of human pre-implantation embryo single-cell transcriptomes reveals the dynamics of lineage specification. *Cell Stem Cell* **28**, 1625-1640. e1626 (2021).

14  McCracken, I. R. *et al.* Transcriptional dynamics of pluripotent stem cell-derived endothelial cell differentiation revealed by single-cell RNA sequencing. *European heart journal* **41**, 1024-1036 (2020).